\documentclass{article}
\usepackage{amsmath,graphicx}
\usepackage{microtype}
\usepackage{booktabs}
\usepackage{float}
\usepackage[dvipsnames]{xcolor}
\usepackage[preprint]{spconf}

\copyrightnotice{\copyright\ IEEE 2021}
\toappear{To appear in {\it Proc.\ ICASSP2021, 6-11 June 2021, Toronto, Ontario, Canada}}

\definecolor{thangcolor}{rgb}{0.635,0.998,0.722}

\definecolor{floriancolor}{rgb}{0.998,0.635,0.722}

\title{META-LEARNING FOR IMPROVING RARE WORD RECOGNITION IN END-TO-END ASR}

\name{Florian Lux and Ngoc Thang Vu}
\address{University of Stuttgart\\Institute for Natural Language Processing\\70569 Stuttgart, Germany}

\begin{document}
\ninept

\maketitle

\begin{abstract}
We propose a new method of generating meaningful embeddings for
speech, changes to four commonly used meta learning approaches to
enable them to perform keyword spotting in continuous signals and an
approach of combining their outcomes into an end-to-end automatic
speech recognition system to improve rare word recognition. We
verify the functionality of each of our three contributions in two
experiments exploring their performance for different amounts of
classes (N-way) and examples per class (k-shot) in a few-shot setting.
We find that the speech embeddings work well and the changes to
the meta learning approaches also clearly enable them to perform
continuous signal spotting. Despite the interface between keyword
spotting and speech recognition being very simple, we are able to
consistently improve word error rate by up to 5\%.
\end{abstract}


\section{INTRODUCTION} 
    While end-to-end (E2E) \cite{graves2014towards} deep learning (DL) models brought great
improvements to the field of automatic speech recognition (ASR)
in recent years and reduced word error rates (WER) on benchmark
datasets significantly \cite{li2020comparison}, they also come with a set of problems.
Even though they are designed to work well with unknown words,
since they usually concatenate subword-units and can thus essentially
produce any textual transcription, they tend to perform even worse on
rare words than classical DL ASR systems \cite{spell2019}. This is due to them
having an internal latent language model, which is biased and cannot
be tweaked. For example, an ASR trained on the LibriSpeech corpus
is very likely to transcribe the name \textit{Hilde} as \textit{Hilda} or the name \textit{Josef}
as \textit{Joseph}.

Even though those words are rare, they tend to be of great importance
for many tasks. A good example for this is automatic meeting
transcription. People frequently address each other by name in a
meeting. And meeting participants can have very unique names with
unique pronunciations. So for the transcriptions, the prior knowledge
which names to expect could be very helpful. Incorporating it
however is a non-trivial task, because E2E approaches cannot be fine-tuned
easily, since they lack a lexicon or a pronunciation dictionary.

Proper nouns have been identified as a challenging problem in
ASR for a while now \cite{proper2014}. Recently some approaches have arisen to
tackle this challenge with E2E ASR using a specialised architecture
and losses  \cite{peyser2020improving} or using specific data and training procedures to better
represent contextual information \cite{alon2019contextual}. Our approach is meant for rare
words in general, however in this work we choose rare proper nouns
as exemplary data and use few-shot learning to improve performance
on them.
     
In meta learning  \cite{schmidhuber1987evolutionary} using the metric space approach (MSML)
\cite{Goldberger2004NeighbourhoodCA}, an embedding function is trained to transform datapoints into a
metric space, where comparisons purely based on positions are possible.
Since structural knowledge about a task is within the embedding
function, new information can be considered by making comparisons
to reference samples of data we want to adapt to. Essentially clustering
is performed in a latent space. MSML could provide solutions
for our goal of improving recognition of rare words in highly specific
contexts, since it can operate with as little as one reference datapoint
of the data we want to adapt to, it doesn’t need any retraining or time
consuming tweaking mechanisms for new reference samples, and the
decisions made on the basis of the comparisons in the metric space
are simple and interpretable and can thus be integrated into an E2E
workflow with few complications.

    We propose a two part ASR system that integrates metric space
representations of expected difficult keywords into the E2E pipeline.
An ASR system is built using the Transformer architecture \cite{vaswani2017attention} which
has been shown to outperform any prior architecture for most speech
related tasks\cite{karita2019comparative, li2019neural, vila2018end, dong2018speech} by heavily relying on attention mechanisms
\cite{bahdanau2015neural}. The intermediate representation that the Transformer-encoder
produces is used as prior embedding and then embedded
further into a metric space on a frame-shifted window basis inspired
by previous work on keyword spotting \cite{chen2014small}. Prior approaches of embedding
audio include using a skip-gram model for speech \cite{chung2018speech2vec} and
using convolutional neural nets (CNNs) for general purpose audio
detection  \cite{audiomax}. By using the Transformer as embedding function, we
hope to get more phonetically rich embeddings. The metric space
embeddings are then compared to reference samples using various
renown MSML approaches such as Siamese Networks \cite{bromley1994signature}, Relation
Networks \cite{sung2018learning}, Prototypical Networks \cite{snell2017prototypical} and Matching Networks
\cite{vinyals2016matching}. The results of these comparisons are taken into account when
decoding the orthographic representation of the utterance.

Our main contributions are as follows: we propose a new method
of generating meaningful embeddings of spoken language, we redesign
MSML approaches to spot keywords in a continuous signal
and we showcase an approach to integrate the results into an E2E
ASR system. We verify our proposed methods on two severely different
datasets and observe the embeddings working well, the keyword
spotting achieving high F$_1$ scores and the keyword spotting ASR
interface improving rare-word WER by 1.6\% on average.

\section{PROPOSED APPROACH}

\begin{figure}[htb]
        \centering
        \centerline{\includegraphics[width=.48\textwidth]{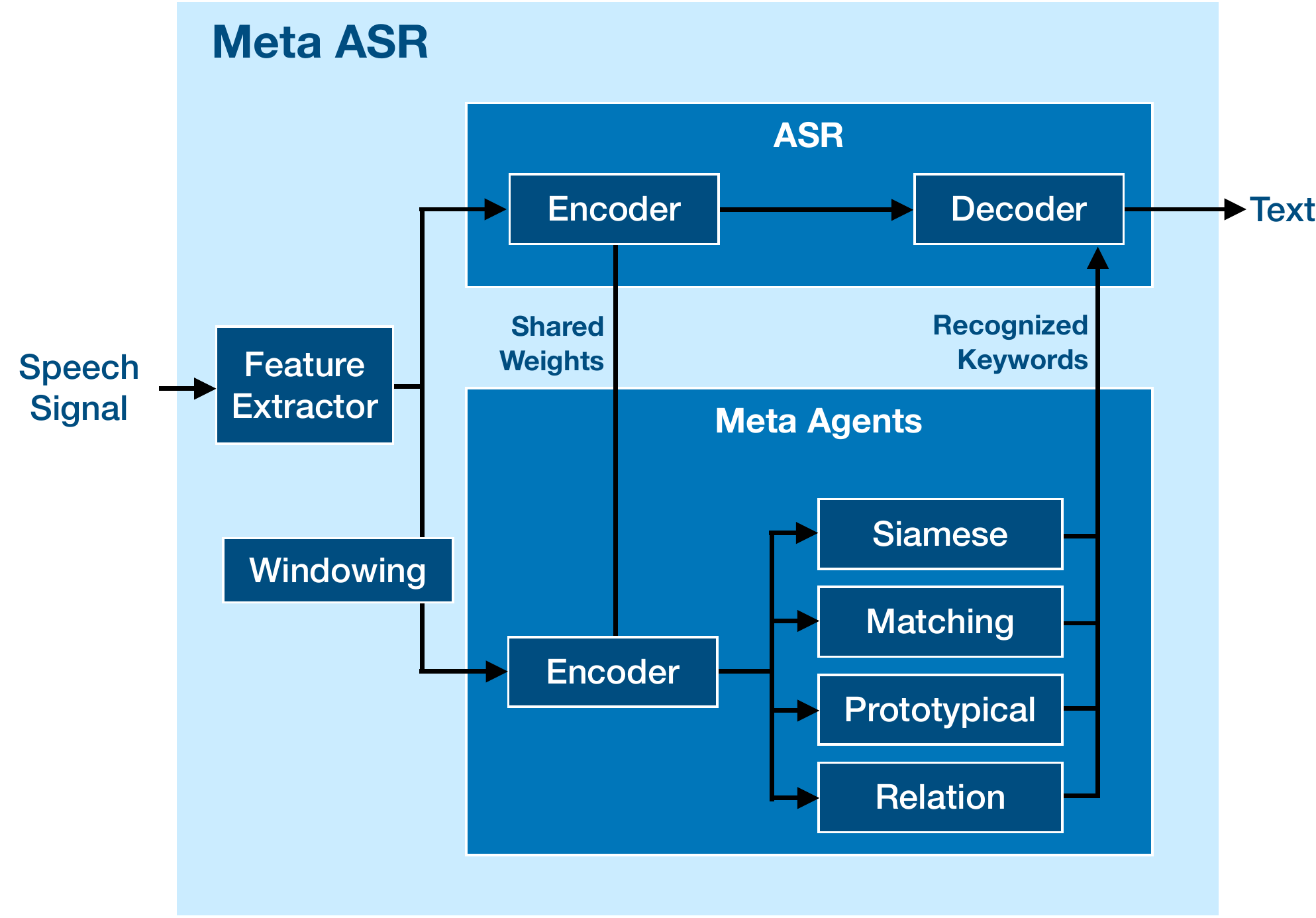}}
        \caption{Overview of how the components are combined}
        \label{sys}
    \end{figure}

An overview of the entire system we built can be seen in figure \ref{sys}.
The ASR encoder feeds into the ASR decoder, and also into each of
the four MSML agents in a windowed fashion. The ASR decoder
produces hypotheses and the MSML agents (usually only the one you
favor) produce a list of recognized keywords. The selection of the
best hypotheses that the decoder’s beamsearch produces takes the list
of recognized keywords into consideration.

    \subsection{Semantics of Audio} 
    
    We need to abstract away from speaker, microphone, recording environment
and so on, but retain the phonetic information. To bypass
this challenge, we use the output of the encoder of a Transformer
ASR as the input to the metric space embedding function. The Transformer
encoder has been trained on thousands of hours of speech
beforehand and thus abstracts away from surface level features and is
robust against data variations already.

    To use the ASR system as embedding function, a feature matrix
derived from an arbitrary audio signal is fed to the encoder of the
Transformer, yielding an embedding matrix with 512 dimensions
per timestep in our case. The Transformer applies attention with
respect to the entire input sequence itself, so in order to get stable
embeddings we have to limit the attention span to the size of our
window. This is done by windowing over the signal first and encoding
one appropriately sized chunk at a time using the ASR.

    \subsection{Meta Learning for Spotting} 
    
    MSML approaches are usually used for classification on clean data,
mostly images. We however want a network that can be applied
efficiently to a sequence of windows of a signal and rather than
classifying the correct class, spot whether something is present or
not. This requires some major changes to the architectures of the four
meta learning approaches used.
    
        All of the approaches are redesigned to look at one class at a
time. Each network assigns a similarity score to a pair of query and
a member of the support set of a class. The query is a window of
ASR-encoded signal. For each class, the highest similarity score that
any of its supports achieves with any of the windows is saved. In
the end, the score of each class is compared against a manually set
threshold and if it peaks above, it is added to the list of recognized
keywords that is passed on to the ASR decoder.

For the Siamese Network, the support and the query are embedded
into the metric space, and the cosine similarity (CosSim,
normalized dot product, measure how well vectors align) of the resulting
vectors is computed. The scoring mechanism is displayed in
equation \ref{eq:s}. An encoded matrix M is denoted as enc(M), the query as
$q$, the support as $s$
        
        \begin{equation}
            \text{score}_\text{class}^{q} = \max_{s \in \text{class}} \cfrac{\text{enc}(q)\cdot\text{enc}(s)}{||\text{enc}(q)||\cdot||\text{enc}(s)||}
            \label{eq:s}
        \end{equation}
        
        In our Relation Network implementation, the query and a support are again encoded using the metric space encoder. Here however the similarity function is learnt, not static. The two vectors are concatenated and passed though a small multi layer perceptron to get the similarity score. The relation scoring mechanism can be seen in \ref{eq:r}, $relat$ denotes the relation function, $concat$ denotes vector-concatenation.
        
        \begin{equation}
            \text{score}_\text{class}^{q} = \max_{s \in \text{class}} \text{relat}(\text{concat}(\text{enc}(q),\text{enc}(s)))
            \label{eq:r}
        \end{equation}
        
        Our Prototypical implementation takes in multiple supports at a time, which are then averaged after they have been encoded to build a prototype of the corresponding class, which is then compared to the query using CosSim. For the case of k=1, it is equivalent to the Siamese Network. The scoring mechanism is displayed in equation \ref{eq:pn}, the calculation of the prototype is displayed in equation \ref{eq:p}.
        
        \begin{align}
            \text{score}_\text{class}^{q} &= \cfrac{\text{enc}(q)\cdot\text{enc}(p_{\text{class}})}{||\text{enc}(q)||\cdot||\text{enc}(p_{\text{class}})||} 
            \label{eq:pn}
            \\
            p_{class} &= \sum_{s \in \text{class}} \cfrac{s}{|\text{class}|}
            \label{eq:p}
        \end{align}

        The Matching Network is usually designed to benefit from fully conditional encoding functions into a metric space, which rely on attention. For this task however, this setup did not converge at all. Instead we use again a simple encoder like the other networks, however with the output being sequences. We apply attention from the query to the support sequence. The conceptual matching lies within the properties of the resulting attention matrix, namely how well the two sequences align. And to get a similarity score from the attention matrix, we pass it through a small long short-term memory layer \cite{hochreiter1997long}. The matching scoring mechanism is displayed in equation \ref{eq:m}, $att$ is an attention mechanism and $reduce$ is any method to reduce over a sequence axis.
        
        \begin{equation}
            \text{score}_\text{class}^{q} = \max_{s \in \text{class}} \text{reduce}(\text{att}(\text{enc}(q),\text{enc}(s)))
            \label{eq:m}
        \end{equation}

    \subsection{Integration with Speech Recognition} 
       
       The decoder of the ASR produces a set of hypotheses using a beamsearch.
One of them is chosen as the final ASR output. This choice
usually just depends on the ranking that the beamsearch itself assigns
to the hypotheses. In our system, this choice is influenced by a list of
recognized keywords that the MSML agents produce. If a keyword is
spotted, hypotheses that contain the word are moved to the front of
the ranking. The spotting at this level is binary.

       To illustrate how this works, assume we have a sentence that
contains the name Noirtier. Shown below are the beginnings of the
hypotheses that our ASR yields for the sentence.

        \begin{itemize}
            \item $<$eos$>$ nautier was near the bed ...
            \item $<$eos$>$ natier was near the bed ...
            \item $<$eos$>$ nartier was near the bed ...
            \item $<$eos$>$ noirtier was near the bed ...
        \end{itemize}

       The correct name appears in the hypotheses, however only as the
fourth option. Now if our keyword spotting system detects that the
sentence contains the name, we pick the most probable hypothesis
that contains the desired transcription, which is the fourth here.
        
\section{DATASETS}
    \subsection{Speech Commands Dataset}  
    \label{data} We train the meta learning agents on the Speech Commands dataset \cite{warden2018speech}. It contains a total of 35 different words with 1000 to 3000 samples each. To further enrich the variety of the dataset, we add 160 more names as keywords with 4 to 6 samples each. Those samples come from three different speech synthesis systems and random data augmentations are applied to them. The additional keywords are carefully selected in order to avoid having names in there that are orthographically different, yet pronounced the same. 
    
    \subsection{LibriSpeech Names} 
    The LibriSpeech corpus \cite{panayotov2015librispeech} is the training set for the Transformer ASR and the development set of the MSML agents. It contains over 1000 hours of read speech. For tuning the meta learning agents, a subset is created from the dev-clean portion. A total of 40 proper nouns are selected, which occur between 4 and 10 times in the whole corpus. For each of them, 4 sentences containing them are selected. From each of the utterances, the name that occurs in them is manually cut out and labelled to create a collection of high quality supports for each name. While we choose proper nouns as exemplary data, what we really want to explore is the performance of rare words, thus the upper limit for their occurrence counts.
    
    \subsection{Singapore English National Speech Names} 
    The Singapore English National Speech Corpus \cite{koh2019building} contains over 3000 hours of speech. The speakers have varying native languages and are recorded with a variety of microphones in different recording environments. We thus believe that the small sub-corpus of sentences that contain proper nouns which we derived from it is well suited to reflect the performance of each of the components in a challenging close-to-real-world scenario. Our derived corpus contains 120 utterances which contain 30 proper nouns and 4 utterances per name. Again, the samples we use as supports are selected and cut manually.
    
\section{EXPERIMENTS}
    \subsection{Setup}
    For the audio embeddings, we train a Transformer ASR on the LibriSpeech corpus using the ESPNet toolkit \cite{watanabe2018espnet, hayashi2020espnet}. The recipe uses 80 Mel-frequency buckets as the input and the training is capped to 40 epochs, similar to the recipe of the IMS-Speech system \cite{denisov2019ims}. The training took 96 hours on 4 Titan X GPUs.
    
    The two experiments use the exact same setup, but look at different aspects. The metric space encoder that is used for all of the MSML agents is built as follows: Two 1D convolutional layers with a kernel size of 3 and 20 filters using hyperbolic tangent as activation function are stacked on top of each other, separated by a maximum pooling layer with a window size of 2. A global max pooling reduces over the sequence axis. The contrastive loss function \cite{lecun2006tutorial, hadsell2006dimensionality} is found to work best. RMSProp is used as the optimizer with a learning rate of 0.001 and a discounting factor of 0.9. 
    
    To train the networks, we sample pairs of audio and assign the label 1 if they are instances of the same word or 0 otherwise. We always alternate between positive and negative examples, even though we expect mostly negative labels during inference. We do so for 800,000 steps, even though convergence shows much earlier, since the MSML networks contain only around 40,000 parameters each.
    
    For the actual experiment we perform a random sampling, since sampling every possible combination of queries and supports is not feasible. This means for any N and k, a random query and k random supports for N classes are selected. Then all four MSML agents are applied to the same randomized setup, as well as the unmodified system as baseline in the second experiment. We get one result for every combination of N, k and agent per run. The random sampling is performed a total of 10 times and the results are averaged.
    
    \subsection{Continuous Signal Spotting} 
    The first experiment aims to explore the impact that k and N have on the performance of the various architectures for spotting keywords in full utterances. The performance is measured in the F$_1$ score, which is the harmonic mean of precision and recall.
        
    \begin{figure}[htb]
        \centering
        \centerline{\includegraphics[width=.5\textwidth]{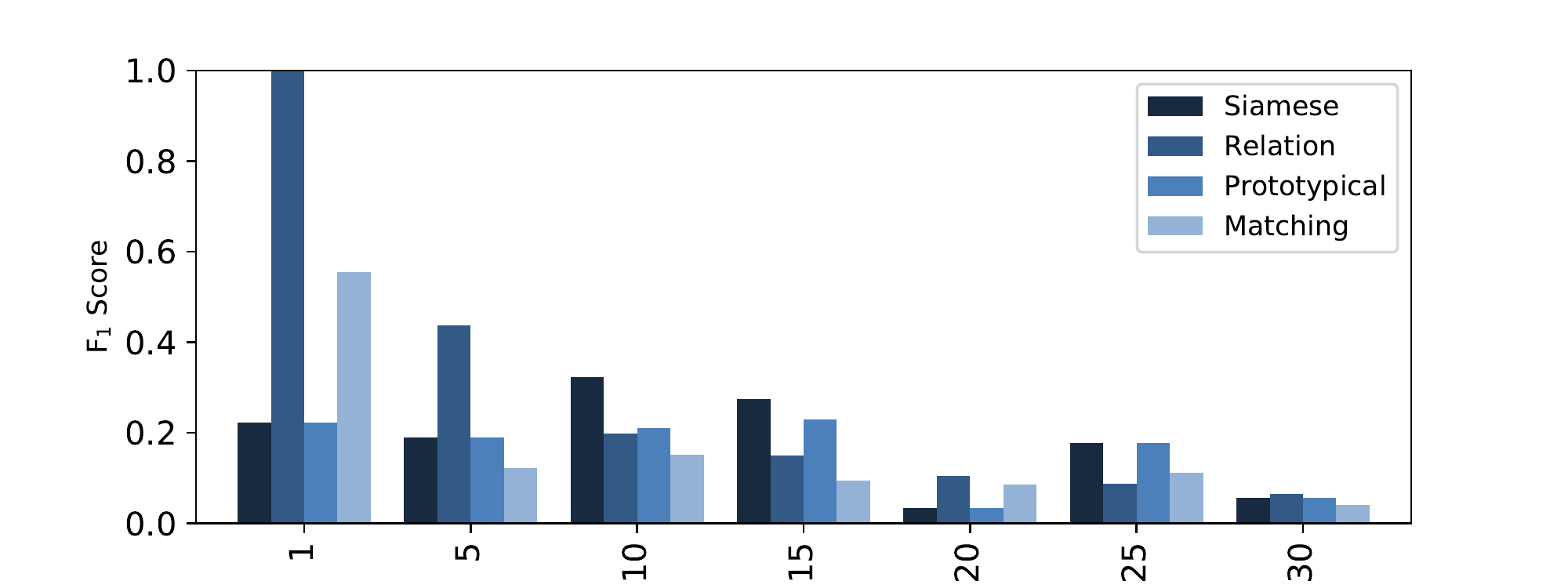}}
        \caption{Spotting F$_1$ scores across different N and agents for k=1}
        \label{k1f}
    \end{figure}
        
    \begin{figure}[htb]
        \centering
        \centerline{\includegraphics[width=.5\textwidth]{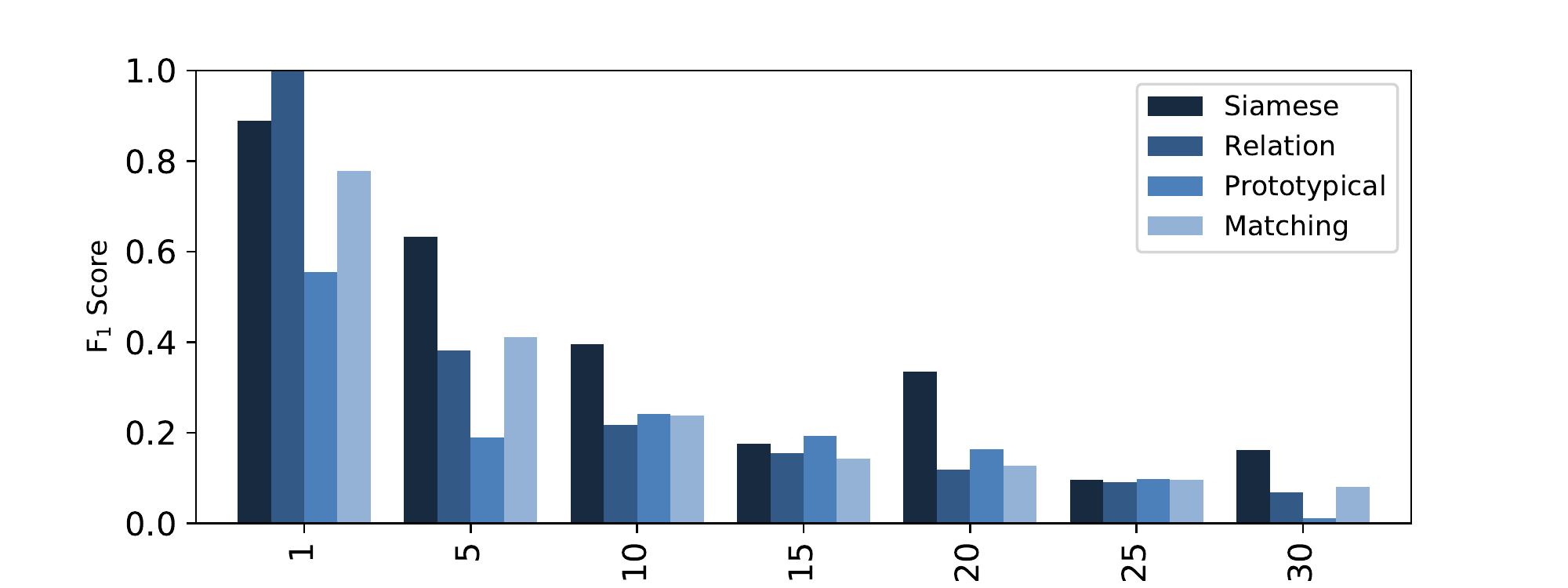}}
        \caption{Spotting F$_1$ scores across different N and agents for k=4}
        \label{k4f}
    \end{figure}
    
    The results on the test set (as seen in figures \ref{k1f} and \ref{k4f}) show Matching and Relation Networks achieving the highest scores. Siamese Networks only take the top spot when given enough supports for each task. The drop in precision and thus in F$_1$ score is severe for an increase in N, however considering that the random baseline decreases from 0.5 down to 0.032, the results are still well above chance. Prototypical Networks do not seem to perform too well overall for this kind of task on the test set, which we suspect is due to some supports disturbing the integrity of the prototype.
    
    \subsection{Speech Recognition Integration} 
    \begin{table}[thb]
        \centering
        \begin{tabular}{l c c c c c c c}
             \toprule
             \hspace{\fill}N = & 1 & 5 & 10 & 15 & 20 & 25 & 30\\
             \midrule
             \multicolumn{7}{l}{\textit{k = 1}} \\
             Siamese       & \textbf{12.2} & \textbf{9.1} & \textbf{10.2}  & \textbf{12.6} & \textbf{9.3} & \textbf{20.1} & \textbf{13.3} \\
             Relation      & \textbf{12.2} & \textbf{9.1} & \textbf{10.2}  & \textbf{12.6} & \textbf{9.3} & \textbf{20.1} & \textbf{13.3} \\
             Proto         & \textbf{12.2} & \textbf{9.1} & \textbf{10.2}  & \textbf{12.6} & \textbf{9.3} & \textbf{20.1} & \textbf{13.3} \\
             Matching      & \textbf{12.2} & \textbf{9.1} & \textbf{10.2}  & \textbf{12.6} & \textbf{9.3} & 21.0 & 14.3 \\
             Vanilla      & 16.4 & 10.9 & 12.4 & \textbf{12.6} & \textbf{9.3} & 21.0 & \textbf{13.3} \\
             \midrule
             \multicolumn{7}{l}{\textit{k = 4}} \\
             Siamese  & \textbf{11.7} & \textbf{8.8} & \textbf{10.1} & 14.2          & \textbf{6.9}  & \textbf{13.2} & \textbf{6.1} \\
             Relation & \textbf{11.7} & \textbf{8.8} & \textbf{10.1} & 14.2          & \textbf{6.9}  & \textbf{13.2} & \textbf{6.1} \\
             Proto    & \textbf{11.7} & \textbf{8.8} & \textbf{10.1} & 14.2          & \textbf{6.9}  & \textbf{13.2} & \textbf{6.1} \\
             Matching & \textbf{11.7} & 9.7         & \textbf{10.1} & \textbf{13.4} & 10.0          & \textbf{13.2} & 8.6\\
             Vanilla & 15.2          & 12.4         & 15.1          & 14.2          & \textbf{6.9}  & 15.3          & \textbf{6.1} \\
             \bottomrule
        \end{tabular}
        \caption{Word Error Rates in \% across N, k and approaches}
        \label{tab:wer}
    \end{table}
    
    The second experiment focuses on the ASR performance and the
impact that the meta learning agents’ results have on it. This is
measured using the WER of the ASR with and without the meta
learning agents as baseline (called Vanilla here). Inconsistencies in
the Vanilla WERs originate from the randomized setup to ensure
comparability of the approaches within each N and k. The WER
in this evaluation is only counting mistakes that originate from the
keywords. All other mistakes are ignored, as the meta learning agents
have no effect on them. The reranking in this setup considers all of
the hypotheses that the beam search of the decoder (beamsize = 4)
produces
    
    As can be seen in table \ref{tab:wer}, the meta learning agents help the
ASR slightly in most cases for any N and k. Interestingly there is
very little difference between the meta learning agents. One more
remarkable thing is that the Matching Networks seem to be the only
cases where the WER actually increases over the vanilla system with
no augmentations. This is probably due to them having low precision,
but very high recall in the first experiment, so a lot of false positives.

\section{DISCUSSION} 
    \noindent  \textbf{Impact of Transformer based Embeddings} To begin we want to address insights into the usefulness of the Transformer encoder as embedding function. The results on the development set and the results on the test set for both experiments are remarkably close, even though the data is severely different. This good performance across corpora hints at the Transformer ASR embeddings working as intended. In comparison, when the same MSML nets are trained on Mel frequency buckets directly, the performance across corpora shows a massive drop from development to test set.
    
\begin{figure}[h]
        \centering
        \centerline{\includegraphics[width=.48\textwidth]{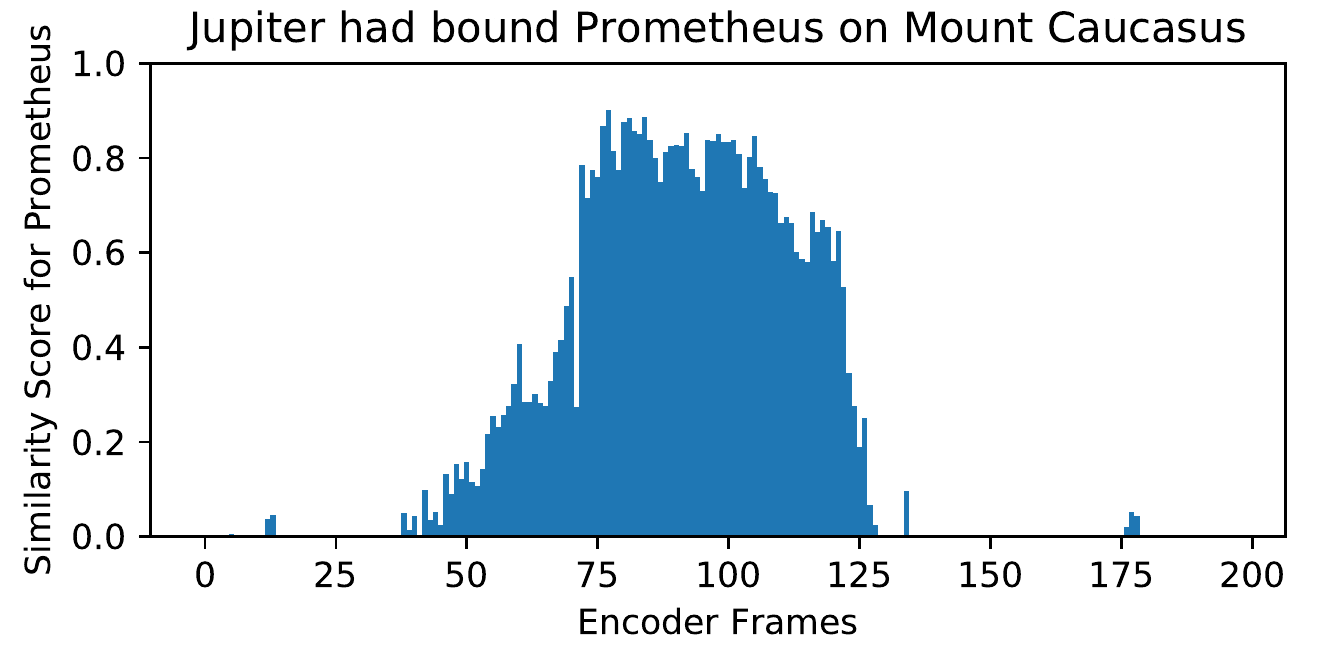}}
        \caption{Siamese similarity score over frames of an utterance given one support for the keyword Prometheus}
        \label{sot}
    \end{figure}

    \noindent  \textbf{The Spotting Process} To provide some insight into the spotting process over a whole signal, figure \ref{sot} shows the Siamese similarity score given a support of the word Prometheus over time. At the point in the sentence where the keyword actually appears, the score peaks over 0.95. In most longer signals however, there are random peaks aside from this main peak. Looking at the transcriptions, unfortunately no specific reason becomes apparent. The margin between an actual peak and a random peak across signals is however big enough for a consistent threshold to be set (we choose 0.8).
    
    \noindent  \textbf{Impact of N, k and Support Quality on MSML Agents} Next we
want to give some insights into cases of MSML agent performance
which we find interesting. The MSML agents outperform the random
baseline by a huge margin for any N and k, with the performance decreasing
for large N. The Relation Network manages to perform well
even with few supports and surprisingly even in cases where the only
support it is given is pronounced slightly wrong. The Prototypical and
Siamese Networks greatly benefit from more supports, however the
prototype is sometimes disrupted by heavily mispronounced names
in the supports. This is not the case in the development set, where the
Siamese and the Prototypical approach actually yield the best results,
together with the Relation approach. So robustness against imperfect
supports appears to be important for close-to-real-world scenarios.
    
    \noindent  \textbf{Impact of imperfect Speech on ASR} One interesting insight about the keyword spotting ASR interface are the cases where the regular ASR tends to make mistakes. Looking into the data and the hypotheses of the ASR, we find that usually the most probable ASR hypothesis matches exactly what the speaker of the utterance articulated. This is however not always the desired transcription, since speakers tend to be sloppy, especially with names, or mispronounce them entirely. 
    
    \noindent  \textbf{Impact of the Keyword Spotting Integration} Another interesting observation can be made with the cases where the hypothesis was not changed, despite the MSML agents correctly spotting a keyword. This happens quite frequently for names that follow phonotactics that are very different from English, since they do not appear in the hypotheses at all and could thus also not be reranked. At the same time this is also a good thing, since especially the Matching Network often spotted keywords which were not present. The hypothesis does not change then, because the word in question simply does not appear in the selection. So the ASR hypotheses act as a filter to the MSML agents and the MSML agents act as a filter to the ASR hypotheses.

\section{Conclusions} 
    In this paper we introduce the idea of using a Transformer ASR encoder as an embedding function, which enables stable good performance across severely different corpora. We also discuss a setup that enables four commonly used MSML approaches to perform keyword recognition on arbitrary reference samples in continuous signals. Finally we introduce a system that combines ASR with keyword spotting to enable modular and simple fine-tuning of neural E2E ASR to expected vocabulary. Experiments show a keyword recognition performance much higher than chance across different setups and an improvement in WER, indicating that the fine-tuning works despite the mechanism being a rather simple proof of concept.

\bibliographystyle{IEEEbib}
\bibliography{refs}

\end{document}